\documentclass[aps,pra,groupedaddress,twocolumn,floatfix,amssymb,noeprint]{revtex4-2}
\usepackage[english]{babel}
\usepackage[pdftex]{graphicx}
\usepackage{dcolumn}

\usepackage[hidelinks]{hyperref}
\usepackage{units}
\usepackage{booktabs}

\usepackage{braket}
\usepackage{amsmath} 
\renewcommand{\selectlanguage}[1]{}
\newcommand{\Er}{\,E_{\rm{r}}}
\newcommand{\al}{\,a_{\rm{\ell}}}

\usepackage{newtxtext,newtxmath}
\usepackage{graphicx}
\usepackage{upgreek}
\usepackage{float}
\usepackage{xcolor}
\usepackage[normalem]{ulem}
\begin{document}
\newcommand{\old}[1]{\footnotesize #1}
\renewcommand\thesubsection{\arabic{subsection}}

\title{Quantum-gas microscopy\ and Talbot interferometry of the Bose-glass phase}

\author{
	Lennart~Koehn$^{1\ast}$,  Christopher Parsonage$^{1\ast}$,      Callum W. Duncan$^{1,2}$,  Peter Kirton$^{1}$, \\
     Andrew J. Daley$^{1,3}$, Timon Hilker$^{1}$,   Elmar Haller$^{1}$,
        Arthur La Rooij$^{1}$,    Stefan Kuhr$^{1}$\\
    \smallskip
    \textit{	\small$^{1}$Department of Physics, University of Strathclyde, Glasgow G4 0NG, UK\\
        \small$^{2}$Aegiq Ltd., Cooper Buildings, Arundel Street, Sheffield S1 2NS, UK.\\
	\small$^{3}$Department of Physics, University of Oxford, Oxford OX1 3PU, UK.\\}
    \smallskip
	\small$^\ast$These authors contributed equally to this work.
}

\begin{abstract}

Disordered potentials fundamentally affect transport and coherence in quantum systems, giving rise to a Bose-glass phase in interacting bosonic systems - an insulating yet compressible phase lacking long-range coherence. Directly measuring a reduced coherence length of the Bose glass has been a outstanding challenge. We address this by employing Talbot interferometry combined with single-atom-resolved detection in a quantum-gas microscope. Using ultracold bosonic atoms in a two-dimensional lattice with site-resolved, reproducible disorder, we identify the Bose-glass phase through in-situ density distributions and particle-number fluctuations, quantified via the Edwards-Anderson parameter, and through the visibility of interference patterns after time-of-flight.
By driving the system across the Bose-glass phase, we further observe signatures of non-ergodic dynamics. Our studies provide a starting point to further explore disordered systems in and out of equilibrium, and are relevant for understanding the dynamics and stability of disordered and glass-like quantum states in solid-state systems.

\end{abstract}

\maketitle

\pagenumbering{arabic}


\noindent
Disordered potentials govern electron transport and thermalization in real materials \cite{krusin1994bose, berthier2011theoretical,ducry2020electro}, and have also been  studied in many-body quantum systems with ultracold atoms  \cite{sanchez2010disordered, shapiro2012cold, pollet2013review, Abanin2019}, in both experiment and theory. For non-interacting quantum systems, destructive interference from scattering in the disordered medium leads to Anderson localization  \cite{Anderson1958}, a phenomenon that has been observed for example in photonic lattices \cite{schwartz2007transport,lahini2008anderson} and in cold-atom systems \cite{roati2008anderson,Pasienski2010,Kondov2011,jendrzejewski2012three, white2020observation}. In general, the addition of interactions gives rise to a complex set of phenomena including, at low energies, the emergence of the Bose glass \cite{Fisher1989,Lee2001,Fallani2007,aleiner2010finite} and spin glasses \cite{Binder1986} in bosonic and fermionic systems.  The Bose glass has been examined using various theoretical approaches. Mean-field theories work less well near phase transitions because of statistical fluctuations, and this also affects disordered systems \cite{lin2011superfluid}. While numerical methods can capture transitions for smaller system sizes in the presence of disorder, they face challenges with finite-size scaling \cite{soyler2011phase}, making the Bose glass an interesting system to study experimentally using a quantum simulator.

The Bose glass is both insulating and compressible \cite{Weichman2008,Pollet2009}, it has a gapless energy spectrum, and, unlike a superfluid, it has no long-range phase coherence, but instead it develops `superfluid puddles' that are not connected to each other. It was initially believed to exist solely as a ground-state phase \cite{Fisher1989,Pollet2009}, however, recent studies suggest that it can also occur at finite temperature \cite{aleiner2010finite,bertoli2018finite,zhu2023thermodynamic}.  
In earlier studies,  excitations in a quantum gas following a sudden change (quench) of optical speckle patterns have been observed \cite{Meldgin2016}, consistent with the existence of the Bose-glass phase, alongside a reduction of coherence \cite{nagler2022observing}. Another important question is how disorder affects ergodicity and thermalization. In quantum many-body systems, establishing thermal states locally requires the propagation of correlations, which is restricted to exponentially long timescales when the wavefunctions are localized, either in an insulating or a disordered system. Such phenomena were observed with fermionic and bosonic atoms \cite{schreiber2015observation,Choi2016,rispoli2019quantum}. Signatures of non-ergodic behavior have also recently been observed in the Bose glass in a two-dimensional optical quasicrystal \cite{yu2024observing}, by probing the adiabaticity of the Bose-glass-to-superfluid transition.

In this study, we probe bosonic atoms in a two dimensional square optical lattice onto which we project a known, controllable, and repeatable disorder light potential at the scale of individual sites \cite{Choi2016}. Single-atom resolved detection gives access to the in-situ distribution and particle fluctuations, allowing us to obtain a local measurement of the Edwards-Anderson parameter \cite{edwards1975,Morrison2008,Thomson2016,Abadal2020Thesis}. We study a regime between the superfluid and Mott-insulating phases in the presence of disorder that shows increased values of the Edwards-Anderson parameter and a reduced visibility in the interference pattern after a time-of-flight expansion. These are complemented by a quantitative measurement of the coherence length using Talbot interferometry. We find that the coherence length   reduces with increasing disorder strength, consistent with the formation of `superfluid puddles'. Finally, we investigate adiabaticity during transitions between the different regions in the phase diagram, observing non-adiabatic behavior when crossing into the disordered phase,  consistent with properties  of the Bose-glass phase.

\begin{figure}[t] \centering{}
    \includegraphics[width = 1.0\linewidth]{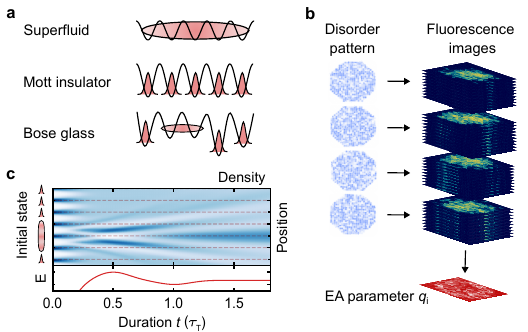}
      \caption{\label{fig:Figure1}\textbf{Characterising phases and coherence in the disordered Bose-Hubbard model.} \textbf{a} Illustrations of the phases of the disordered Bose-Hubbard model highlighting the degree of coherence between lattice sites. \textbf{b} Schematic to calculate the Edwards-Anderson (EA) parameter. We record ten fluorescence images for a number of disorder patterns at the same disorder strength and average the occupation on each lattice site to obtain disorder averages, see Eq.\,(\ref{EAparamter}). \textbf{c} We probe the coherence length of the system along one axis using an interferometry technique based on the Talbot effect. Top panel: Time evolution of the sketched one-dimensional initial state after switching off the horizontal lattice. Initially phase coherent regions show (shifted) revivals after (half) integer Talbot times, $\tau_{\rm T}$. Horizontal dashed lines indicate the positions of the lattice sites. Bottom panel: Total energy after recapturing the atoms in the original lattice after a duration $t$. The decay time of the resulting oscillations can be related to the coherence length of the initial state (Appendix).} 
\end{figure}

\begin{figure*}[!t] 
   \includegraphics[width = 1.0\linewidth]{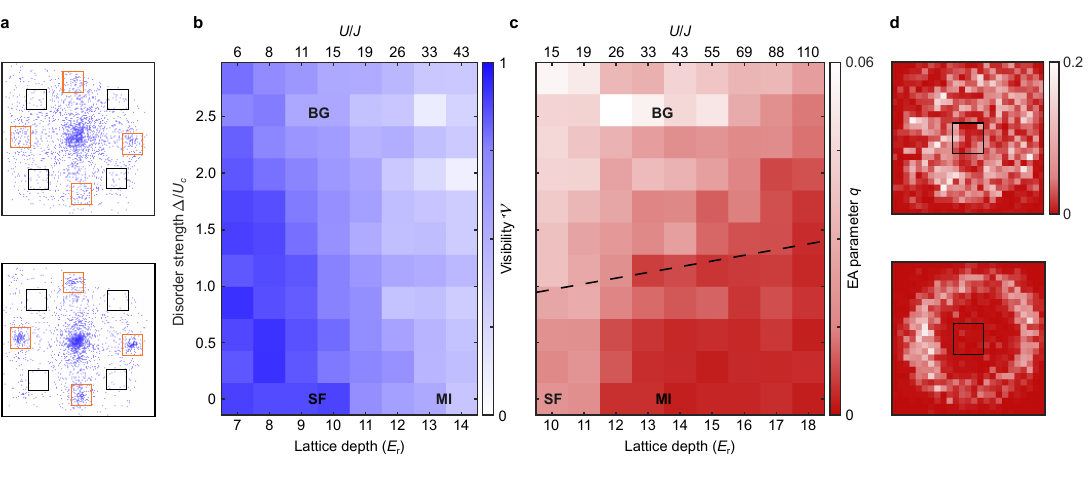}%
   \vspace{-0.5cm}
   \caption{\label{fig:Figure2}\textbf{Probing the disordered Bose-Hubbard model.} \textbf{a} Average occupations of ten time-of-flight images at $U/J = 15$ and $\Delta/U_c = 2.8$ (top), $\Delta/U_c = 0$ (bottom), where $U_c$ is the interaction energy at $(U/J)_c$, plotted on a logarithmic color scale for clarity. Orange and black boxes highlight the regions used to determine $n_{\rm max}$ and $n_{\rm min}$, respectively, used to calculate the visibility, $\mathcal {V}$. \textbf{b} Visibility as a function of lattice depth and disorder strength. Our analysis identifies superfluid (SF), Mott insulating (MI) and Bose glass (BG) phases. \textbf{c} Edwards-Anderson (EA) parameter (Eq.\,\ref{EAparamter}) averaged over the central $5 \times 5$ lattice sites (black box in bottom panel of \textbf{d}). The dashed line indicates $\Delta/U=1$, above which the disorder strength is larger than the energy gap in the ideal Mott insulator \cite{Pollet2009}. \textbf{d} Site-resolved EA parameter, $q_i$, at $U/J = 55$, using four different disorder realizations with ten in-situ images each, for $\Delta/U_c = 2.8$ (top) and $\Delta/U_c = 0$ (bottom). Error bars of $\mathcal {V}$ and $q$ are shown in Figs.~\ref{fig:SI_extendedDataFig2_QMC} and~\ref{fig:SI_ExtendedData_Fig2_EA} (Supplemental Material).}
\end{figure*}

\subsection{Bose-Hubbard model with disorder}
\label{sec:BoseGlass}

The disordered Bose-Hubbard model is described by the Hamiltonian
\begin{equation}
    \hat{H} = -J \sum_{\langle i,j\rangle} \hat{a}_i^\dagger \hat{a}_j + \sum_{i} \frac{U}{2} \hat{n}_i(\hat{n}_i-1) + \sum_{i}(\epsilon_i - \mu)\hat{n}_i + \sum_{i}\Delta_i \hat{n}_i \,,
\label{eq:BHM}
\end{equation}
where $J$ is the tunneling strength between neighboring lattice sites, $\hat{a}_i^\dagger$ and $\hat{a}_j$ are the bosonic creation and annihilation operators on neighboring sites $i$ and $j$, respectively, $U$ is the on-site interaction energy, $\hat{n}_i=\hat{a}_i^\dagger\hat{a}_i$ is the number operator,  $\epsilon_i$ is the local energy shift from the harmonic confinement, and $\mu$ is the chemical potential. $\Delta_i$ is a local potential that creates random energy offsets on each lattice site, with a uniform box distribution $\Delta_i \in [0, \Delta]$, where $\Delta$ is the disorder strength. The model can give rise to a superfluid, a Mott-insulating and a Bose-glass phase (see illustration in Fig.~\ref{fig:Figure1}a). In the strongly interacting regime, $U \gg J$, a Mott-insulating phase exists for weak disorder $\Delta < U$. When the disorder is larger than the energy gap, $\Delta > E_g$, the state becomes gapless \cite{Fisher1989}. In the weakly interacting regime, the disorder results in localization of atoms and a shorter-range phase coherence compared to a superfluid. In a two-dimensional infinite system it is expected that the Bose-glass phase exists between the Mott insulator and superfluid for any non-zero value of $\Delta$ \cite{Pollet2009}.

\subsection{Experimental details}

In our experiments \cite{DiCarli2024}, a cloud of ultracold \textsuperscript{87}Rb atoms is held in a single antinode of a vertical optical lattice (wavelength $\lambda=1064$ nm and depth $V_z=25\Er$, where $\Er=\hbar^2/2m\lambda^2$ is the recoil energy and $m$ is the mass of a \textsuperscript{87}Rb atom). After evaporative cooling, a superfluid of $\sim 200$  atoms is prepared in shallow horizontal lattices ($V_x = V_y \leq 10\Er$). We then vary $U/J$ by changing $V_x$ and $V_y$ along the horizontal axes. We use fluorescence imaging to obtain the in-situ site occupations of the cloud, which has a diameter of $\sim 16$ sites. Typical temperatures are $k_B T \sim 0.1\,U$, measured using radial fits of these occupations in a Mott-insulating state \cite{Sherson2010}.

To create disordered systems, we project a repulsive light potential generated by a blue-detuned laser incident on a digital micromirror device (DMD). At each lattice site, this potential (pictured in Fig.~\ref{fig:Figure1}b) has a random amplitude, $\Delta_i \in [0, \Delta]$, that follows a uniform box distribution (Appendix). Our method of creating disorder patterns is in contrast to most other ultra-cold atom experiments on disordered systems, which typically use speckle patterns, quasirandom, or quasicrystalline potentials \cite{billy2008direct,  Pasienski2010, sbroscia2020observing}. Using the DMD, we can generate not only reproducible disorder patterns, but also maintain phase alignment with the horizontal lattices, which would otherwise drift over time \cite{DiCarli2024}. 

\subsection{Time-of-flight visibility}

In the weakly interacting  regime,  the presence of disorder should result in a reduced coherence, 
which we initially measure via the visibility after time-of-flight \cite{Bakr2010}. 
This allows us to map out the parameter space for different disorder strengths and lattice depths, within a reasonable  experimental runtime. We later complement this with a quantitative measurement of the coherence length  via Talbot interferometry (Section 5) for selected parameters. 

In a time-of-flight measurement, the long-range phase coherence of the superfluid results in peaks in the interference pattern along the horizontal lattice directions. We quantify the visibility of the interference peaks,
  $\mathcal{V} =
  \left(n_{\rm max} - n_{\rm min}\right)/
  \left(n_{\rm max} + n_{\rm min}\right)$, 
by comparing the number of atoms in the peaks ($n_{\rm max}$, orange boxes in Fig.~\ref{fig:Figure2}a) to those in regions diagonal from and equidistant to the central peak ($n_{\rm min}$,   black boxes)~\cite{Gerbier2005}. 
For $\Delta = 0$, the visibility drops for increasing $U/J$ (see bottom row of Fig.~\ref{fig:Figure2}b), indicating a reduction of global phase coherence as a Mott-insulating state starts to form in the center of the cloud.
Interestingly, when we increase $\Delta$, we also observe a decrease in $\mathcal{V}$ at $U/J$ below the superfluid-to-Mott-insulator transition, which occurs at $(U/J)_c = 16.7$ \cite{capogrosso2008monte,soyler2011phase}. 
In the absence of thermal atoms, this reduction in $\mathcal{V}$ is consistent with the formation of the Bose glass. Our observation of a reduced $\mathcal{V}$ with increasing $\Delta$ cannot be attributed to  trivial thermal excitations, as we demonstrate in Section~\ref{sec:Ramps}. There is no clear transition point in our system due to the underlying harmonic confinement, resulting in a spatially varying chemical potential, and finite-size effects. We also compare the visibility measurements to results from Quantum Monte-Carlo simulations (See Supplemental Material), which show the same general trend. 

\subsection{Edwards-Anderson parameter}

The visibility is not a reliable observable for distinguishing the Mott insulator from the Bose glass, as both lack long-range phase coherence (rightmost columns of Fig.~\ref{fig:Figure2}b). Therefore, we employ the Edwards-Anderson parameter, $q_i$, which uses the atom distribution of the in-situ images \cite{Thomson2016,Abadal2020Thesis},
\begin{equation}
    q_i = \overline{\braket{\hat n_i}^2} - \overline{\braket{\hat n_i}}^2\\.
    \label{EAparamter}
\end{equation}
Here, the brackets denote the thermal or ensemble average for a single disorder pattern, while the overline represents an average over different disorder patterns. Specifically, our data sets are divided into four subsets, one for each disorder pattern (Fig.~\ref{fig:Figure1}b). For each lattice site $i$, we determine $q_i$, which is zero when the occupations resulting from each disorder pattern are equal, and positive when the occupations are different from each other. For consistency, the $\Delta = 0$ data were also divided into four subsets, and the small but finite values of $q_i$ reflect thermal fluctuations and atom-number variations between these subsets rather than genuine disorder effects.

For $\Delta = 0$, our measurements show $q_i \approx 0$ in the center of the system (bottom panel of Fig.~\ref{fig:Figure2}d), limited by thermal excitations at finite temperature, indicative of the Mott insulator. We observe a circular region of non-zero values of $q_i$ around the edge of the cloud. This is due to the entropy redistribution to the edge of the cloud resulting from the spatially varying chemical potential, and fluctuations in the total atom number between repetitions of the experiment. 
As we increase $\Delta$, the Mott-insulating region decreases until $q_i$ starts to rise in the center, indicating that no Mott-insulating component remains. To capture the transition from Mott insulator to Bose glass, we average $q_i$ over the central $5 \times 5$ lattice sites, $q$, for increasing $\Delta$ (Fig.~\ref{fig:Figure2}c). When $\Delta>U$, the Bose glass is expected to form across the whole system \cite{Pollet2009}. This predicted behaviour is closely matched by our experimental data, where $q$ increases at disorder strengths $\Delta/U > 1$ (black dashed line in Fig.~\ref{fig:Figure2}c). As we  analyze only the center of the system, where the effect of the harmonic confinement is negligible, we see a clearer transition here compared to the global measurement of the visibility (Fig.~\ref{fig:Figure2}b). We observe $q\approx0.06$ for our maximum $\Delta$, although the maximum possible value is $q=0.25$. Theoretical studies \cite{Thomson2016} have shown that at a finite temperature close to ours ($k_B T \sim 0.1\,U$), the Edwards-Anderson parameter can still distinguish well between glassy phases and thermal fluctuations. However, its magnitude is significantly reduced from $q=0.25$ \cite{Thomson2016,Pal2019}, consistent with our experimental observations.

The Edwards-Anderson parameter is unsuitable to distinguish between the superfluid and Bose glass, see leftmost columns of Fig.~\ref{fig:Figure2}c, for lattice depths lower than $12\Er$ ($U/J = 25.5$) where a superfluid component is present. This disordered superfluid phase has an inhomogeneous density, resulting in a non-zero $q_i$, making it indistinguishable from the Bose glass. Therefore, our measurements of the visibility and of the Edwards-Anderson parameter complement each other. Together, they reveal a region in the phase diagram with properties consistent with the Bose glass. In the measurements that follow, we study the coherence of this region in more detail.

\subsection{Talbot interferometry}
A distinct feature of the Bose glass is the absence of long-range phase coherence, while maintaining short-range coherence over a small number of sites. To measure the coherence length, we employ an interferometry technique based on the Talbot effect, following an earlier experimental demonstration \cite{Santra2017_Talbot}. It involves briefly switching off the lattice potential in one horizontal axis for a duration $t$, before rapidly turning it back on to project the matter wave onto the lattice. At integer multiples of the Talbot time, $\tau_{\rm T}=2m\!\al^2/h=123\,\upmu$s, where $\al$ is the lattice spacing and $h$ is Planck's constant, an initially coherent matter wave interferes constructively at the positions of the lattice sites. At non-integer multiples of $\tau_{\rm T}$, the positions of the interfering atoms are no longer aligned with the lattice sites, causing a sudden increase in potential energy when the lattice is turned back on. After a thermalization time, any potential energy manifests itself as an increase in the cloud temperature, resulting in periodic minima of the cloud width at integer multiples of $\tau_{\rm T}$ and periodic maxima at half-integer multiples (see sketch in Fig.~1c).

\begin{figure}[t] \centering{}
 \includegraphics[width = 1.0\linewidth]{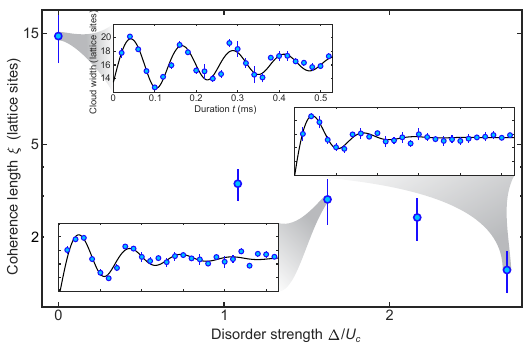}%
     \caption{\label{fig:Figure3}\textbf{Talbot interferometry.}  Coherence length, $\xi$, in units of the lattice spacing, $\al$,  for increasing disorder strength at constant lattice depth ($V_{x,y} = 9 \Er$, $U/J = 11$).
    The coherence length is extracted from the fitted decay time of the Talbot interferometry measurements (insets) by comparing to a numerical calculation (Appendix).
    Error bars represent the $68\%$ confidence intervals of the fitted decay times.    
    Each data point of the curves in the insets results from the average  of five images, each containing $\sim200$ atoms. The error bars are the standard error. Black lines are fits with a damped sine, the decay time of which is used to extract $\xi$ (Appendix). 
    }
\end{figure}

\begin{figure*}[t] \centering
\includegraphics[width = 0.66\linewidth]{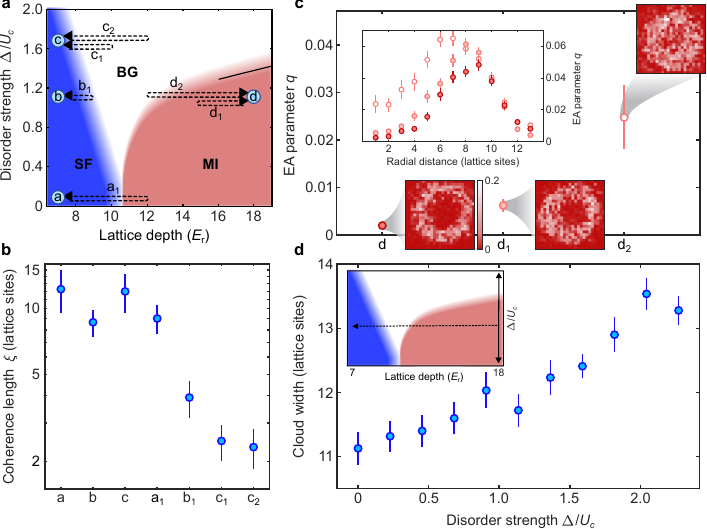}%
   \caption{ \label{fig:Figure4}\textbf{Ergodicity and adiabaticity.} \textbf{a} Approximate phase diagram of the disordered Bose-Hubbard model interpreted from the datasets in Fig. \ref{fig:Figure2}. The blue shaded area highlights the region where the entire cloud is in the superfluid phase and the red shaded region shows where the center of the cloud is Mott-insulating. The blue circles show the points and the arrows the trajectories used for the data sets shown in \textbf{b} and \textbf{c}. \textbf{b} Coherence length, $\xi$, measured using Talbot interferometry for the points and trajectories shown in \textbf{a} (details see text). Each data point results from fits of four repetitions with the same parameters. \textbf{c} EA parameter, $q$, for three points and trajectories as shown in \textbf{a}. Each measurement is an average over four disorder patterns with ten images each. \textbf{d}
    Cloud width after one Talbot time, $\tau_{\rm T}$, after a ramp from the Mott-insulator regime to superfluid regime for increasing disorder strength (see inset where the dashed arrow indicates the trajectory for a data point at a particular disorder strength). 
        }
\end{figure*}

We employ this method to quantitatively probe how the coherence length, $\xi$, of an initially superfluid system is affected by the addition of disorder.
To measure the cloud width, we calculate the average distance of each atom from the center, after a thermalization time of $\unit[250]{ms}$ at $V_{x,y} =6\Er$. The disorder potential is switched off simultaneously with the lattice  and remains off during thermalization and imaging. 
The resulting curve shows a damped oscillation as a function of $t$, (insets in Fig.~\ref{fig:Figure3}), where the presence of $n$ half-periods of oscillation indicates phase coherence over $n$ lattice sites \cite{Santra2017_Talbot}. 
For $\Delta=0$, the oscillation shows a weak decay with all peaks pronounced, indicating long-range phase coherence. When adding disorder, we observe that the oscillations decay more rapidly. At the maximum experimentally accessible disorder strength, only the first two peaks are distinguishable from zero. 
To quantify the coherence length, we fit the data with an exponentially damped oscillation (solid lines in the insets of Fig.~\ref{fig:Figure3}). We compare the fitted decay times to a numerical calculation, which use a density matrix, $\rho(i,j)\sim\exp{(-\lvert i-j \rvert/\xi)}$, with exponentially decaying coherence, see Appendix.
While for $\Delta=0$ we observe a coherence length $\xi=15(3) \al$, the coherence length decreases to $\xi=1.4(3) \al$ for our maximum available disorder strength, see Fig.~\ref{fig:Figure3}. This observation is consistent with the characteristics of a Bose glass, which is expected to exhibit coherence locally in small regions but not across the entire system \cite{Fisher1989,PhysRevB.80.214519}.  Our analysis shows the advantage of the Talbot interferometry measurements to give a quantitative measurement of the coherence length, in contrast to the time-of-flight visibility (Fig.~\ref{fig:Figure2}). In the Mott-insulating phase, the oscillations from the Talbot interferometry measurements are significantly more damped (Supplemental Material) compared to those shown in Fig.~\ref{fig:Figure3}, indicating a coherence length smaller than a lattice site.

\subsection{Ergodicity, adiabaticity, and coherence}
\label{sec:Ramps}
In a finite-size system, a phase transition can be crossed without significant heating if both phases are ergodic. Signatures of non-ergodicity in the Bose-glass-to-superfluid transition were recently observed using a quasicrystalline potential \cite{yu2024observing}. Here, we explore in a square lattice how the Mott-insulator-to-superfluid transition is affected by the presence of disorder. Using the data shown in Figs.~\ref{fig:Figure2}b and c, we draw an approximate phase diagram for our system (Fig.~\ref{fig:Figure4}a) to serve as a guide for the subsequent measurements.

Initially, we perform Talbot interferometry for a set of points in the superfluid phase ($U/J=5.5$), for three different disorder strengths. We find that the addition of the disorder potential does not result in significant heating by holding a disordered superfluid for $400$ ms (points $b$ and $c$ in Fig.~\ref{fig:Figure4}a) and observe long-range coherence comparable to the case without disorder (point $a$).  Importantly, this relates back to our earlier measurements, where the reduced visibility at large disorder strengths (Fig.~\ref{fig:Figure2}b), as well as the quantitative observation of the reduction of coherence length in the Talbot measurements (Fig.~\ref{fig:Figure3}), can then be interpreted as signatures of the emerging Bose glass.

To probe adiabaticity, we first verify that transitioning to the Mott insulator and back to the superfluid at $\Delta = 0$ (trajectory $a_1$ in Fig.~\ref{fig:Figure4}a) restores a comparable coherence length, shown in Fig.~\ref{fig:Figure4}b. This was done by increasing $V_x$ and $V_y$ to reach $U/J=25.5$ in $\unit[200]{ms}$ and vice versa. 
Next, we investigate how entering the region where we identified the Bose glass affects the system's ability to restore phase coherence. Starting with a superfluid with weak disorder and increasing $V_x$ and $V_y$ to reach $U/J=10.8$ in $\unit[200]{ms}$ and vice versa (trajectory $b_1$ in Fig.~\ref{fig:Figure4}a) we observe a significantly reduced coherence length. We conduct two additional measurements, starting from a superfluid with stronger initial disorder to reach $U/J = 14.6$ and $U/J = 25.5$ (trajectories $c_1$ and $c_2$ in Fig.~\ref{fig:Figure4}a). These measurements show a further reduction in coherence length, shown in Fig.~\ref{fig:Figure4}b, consistent with a larger proportion of the cloud entering the Bose glass compared to $b_1$, indicating the inability of the Bose-glass component to adiabatically transition back into the superfluid. 

The observed loss of coherence cannot be attributed to intensity noise on the lattice lasers as we were able to  restore the initial coherence when transitioning from the Mott insulator to the superfluid (trajectory $a_1$). We have also verified that the coherence does not change when extending the duration over which we change $V_x$ and $V_y$ for the presented trajectories from $\unit[200]{ms}$ to $\unit[500]{ms}$, which is the longest accessible timescale in our experiment.

Next, we explore the transition from the identified Bose-glass region to the Mott insulator. We first measure the Edwards-Anderson parameter in a Mott insulator in the presence of weak disorder ($U/J=110$, $\Delta/U_c=1.1$,  point $d$ in Fig.~\ref{fig:Figure4}a), and find $q\approx 0$ in the center of the cloud, as shown in Fig.~\ref{fig:Figure4}c. When we decrease the lattice depth to reach $U/J=55$ and return to $U/J=110$ (trajectory $d_1$), we observe an increase of $q$ in the outer region of the cloud, which we attribute to localization occurring here first due to the lower chemical potential caused by the harmonic confinement. We then perform a similar measurement, changing instead to $U/J=25.5$ and back (trajectory $d_2$), to cross fully into the Bose-glass regime. When attempting to transition back to the original state, we observe an increase of $q$ over the entire cloud, see Fig.~\ref{fig:Figure4}c, again indicating non-adiabaticity. 

Lastly, we investigate the effect of crossing the Bose glass at increasing disorder strengths, starting in the Mott insulator and finishing in the superfluid, see inset in Fig.~\ref{fig:Figure4}d. Specifically, we prepare the system at $U/J=110$ and transition to $U/J=5.5$ within \unit[200]{ms}, before using Talbot interferometry to measure the cloud width after $\tau_{\rm T}$, at the first minimum of the oscillation. We observe an increase in cloud width for increasing disorder strengths, shown in Fig.~\ref{fig:Figure4}d, indicating a loss of coherence. We attribute this to excitations in the system due to the expected non-ergodicity \cite{yu2024observing} when crossing an increasingly large Bose-glass region in the phase diagram.

\subsection{Conclusion}
\label{sec:conclusion}
In summary, we identified the Bose-glass phase experimentally in both weakly and strongly interacting regimes in a two-dimensional square lattice with controllable and reproducible disorder potentials.  Using Talbot interferometry across the superfluid-to-Bose-glass transition, we observed a change from long to short-range coherence.  Additionally, we found that transitioning in and out of the Bose glass is non-adiabatic on the accessible experimental timescales. A similar behavior was also observed in a quasicrystalline lattice at the superfluid-to-Bose-glass transition \cite{yu2024observing}, which opens opportunities to compare and contrast the behavior of these systems and random disorder on Bravais lattices. It may be possible to cross transitions adiabatically on longer timescales, by reducing trap losses and heating, via ultra-low vacuum pressure \cite{Manetsch2024}, and optical lattices with reduced intensity noise. This would enable the exploration of ergodic and non-ergodic phases in a single model, for both weak and strong interactions, through scaling of the adiabatic timescales for traversing transitions or restoring system properties.

Our low-energy studies, combined with Talbot interferometry provide a starting point to further explore disordered systems in and out of equilibrium, and particularly to probe the timescales to establish phase coherence in these systems. This is of relevance to understanding disordered and glass-like quantum states which have been observed in, e.g., cavity polaritons \cite{malpuech2007bose}, quantum magnets \cite{yu2012bose} and overdoped cuprates \cite{tromp2023puddle}.
Future studies with ultracold atoms could focus on the influence of disorder in different lattice geometries \cite{gonzalez2019localisation,mao2020disorder} and lower dimensions \cite{carrasquilla2010characterization,ristivojevic2014superfluid}, or further explore the stability of many-body localization \cite{Choi2016} and the onset of quantum avalanches \cite{leonard2023probing}.
Furthermore, controlled disorder is important in studying frustration and stability of the spin-glass phase in the Fermi-Hubbard model \cite{tusch1993,Sanpera2004spinglass,Ahufinger2005spinglass,paredes2005}.
By extending the system sizes is  possible to study the similarities and differences between disordered and quasicrystalline systems further, e.g., by realization of the 2D extension of the Aubry-Andre model \cite{szabo2020mixed,duncan2024critical}, which could feature mobility edges, critical states and give further insights into the nature of many-body localization and the Bose glass \cite{yu2024observing}.

After completing this work, we became aware of a new preprint studying the disordered Bose-Hubbard model using a superconducting processor \cite{ticea2025}. 

\paragraph*{Acknowledgments}
We acknowledge support by the Engineering and Physical Sciences Research Council (EPSRC) through the Quantum Technology Hub in Quantum Computing and Simulation (QCS) [grant number EP/T001062/1], the Hub for Quantum Computing via Integrated and Interconnected Implementations (QCI3) 
[EP/Z53318X/1], the Programme Grant 'Quantum Advantage in Quantitative Quantum Simulation' (QQQS) [EP/Y01510X/1], and the 2020-2021 Doctoral Training Partnership [EP/T517811/1].

\paragraph*{Author contributions}
LK and CP conducted the experiments and analyzed the data, with support from TH, EH, ALR and SK. CD, PK and AD developed the Quantum Monte-Carlo simulations, while TH led the simulations of the Talbot interferometry. All authors contributed to interpreting the results and writing the manuscript.

\paragraph*{Competing interests}
There are no competing interests to declare.

\paragraph*{Data availability}
The data used in this publication is openly available at the University of
Strathclyde's KnowledgeBase~\cite{Data}.

\newpage

\section*{Appendix} 
\label{sec:Methods}
\subsection*{System preparation and detection}
We use the same quantum-gas microscope setup as detailed in our previous work \cite{DiCarli2024}. Initially, we prepare a Bose-Einstein condensate of $\sim200$ atoms in a single antinode of an optical lattice (wavelength $\lambda = \unit[1064]{nm}$) in the vertical direction, with a depth of $V_z=25\Er$.
To prepare a disordered system in the weakly interacting regime, we turn on the horizontal lattices, $V_x$ and $V_y$, to the desired depth, $V_f \leq 10\Er$, using an s-shaped ramp within 200\,ms and then turn on the disorder potential within \unit[200]{ms} again using an s-shaped ramp. To prepare a disordered system in the strongly interacting regime, we first turn on the horizontal lattices to reach $V_{x,y} = 10\Er$, before adding the disorder while simultaneously increasing $V_{x}$ and $V_{y}$ to their final values, $V_f > 10\Er$ in a further $\unit[200]{ms}$. 
We detect the atoms by collecting their fluorescence light with a high-NA microscope objective and use the Lucy-Richardson deconvolution method \cite{larooij2022} to infer the site occupations.

\subsection*{Generation of disorder patterns}
We create the repulsive disorder potentials using a laser ($\lambda = \unit[666]{nm}$) incident on a digital micromirror device (DMD), similar to our previous work \cite{DiCarli2024}.
The light from the DMD is overlapped with the optical path used for fluorescence imaging of the atoms and projected through the microscope onto the atoms in the 2D plane. The magnification is such that approximately $18 \times 18$ pixels on the DMD correspond to one lattice site. We create a potential at each lattice site (pictured in Fig.~\ref{fig:Figure1}b), with random amplitude $\Delta_i \in [0, \Delta]$, that follows a uniform box distribution. The disorder pattern covers a region with a diameter of 25 lattice sites. 

A motorized flip mirror can direct the light pattern from the DMD onto a monitor camera. We employ an iterative intensity feedback algorithm to shape the Gaussian beam incident on the DMD into the target disorder pattern. The light pattern is blurred by the microscope objective when it is projected onto the atoms. To compensate for this, we convolve the monitor camera images with the point-spread function of the objective to effectively optimize the potential as seen by the atoms.
Once the DMD pattern has been calculated, it is saved and displayed continuously. The disorder strength is controlled solely by changing the laser power. As detailed in our previous work \cite{DiCarli2024}, we measure the phase drift of the optical lattice between experimental realizations by analyzing the position of single atoms in the fluorescence image. This information is then used to shift the position of the DMD pattern for the next measurement, ensuring that the disorder potential remains aligned with the optical lattice. The estimated  phase drift from one experimental realization to the next is $<0.05\al$.

\subsection*{Talbot data analysis and simulation}
\begin{figure}[!b]
    \centering
    \includegraphics[width = 1.0\linewidth]{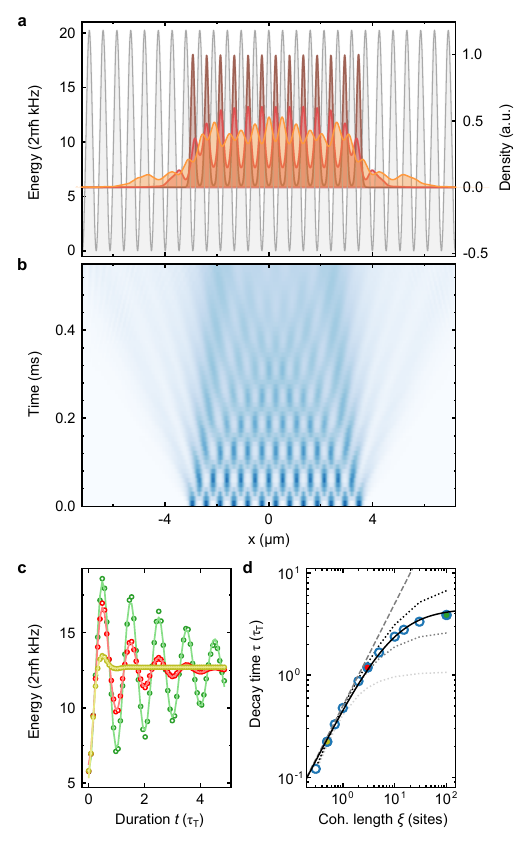}
      \caption[Numerical simulation of Talbot interference]{
      \label{fig:SimTalbot}\textbf{Numerical simulation of Talbot interference.} \textbf{a} Density distribution after free time evolution for $t=[0,1,2.5]\ \tau_{\rm{T}}$ starting from a homogeneous lowest-band distribution over $L=13$ sites with coherence length $\xi=3\al$. \textbf{b} The full time evolution of this density distribution shows (shifted) revivals after (half-) integer Talbot times with decreasing contrast. \textbf{c} Total energy after recapturing the atoms with the lattice after duration $t$. The case from a/b (red)  with  $\xi=3\al$ is compared to initial conditions with $\xi=0.5\al$ (yellow) and $100\al$ (green). The solid lines are fits with exponentially damped harmonic oscillations. \textbf{d} Relation between coherence length $\xi$ and damping of the Talbot oscillations $\tau$ from numerics for $L=13$ (blue points, solid fillings indicate curves in \textbf{c}). For low $\xi$, we find the expected $\tau=\xi/2 \times \tau_{\rm{T}} / \al$ (dashed), while at large $\xi$ the Talbot signal is dominated by finite-size effects (solid line, Eq.\ref{eq:tau_xi}). Dashed lines show results for $L=20,10,5$ (decreasing darkness).   
 }
 \label{fig:SI_TalbotSimulations}
\end{figure}
To analyze the Talbot measurements, we obtain the cloud energy via the cloud width after a thermalization time of 250 ms at $V_{x,y} = 6 \Er$. We measure the cloud width by averaging the distances of each atom from the center of the cloud after thermalization in two dimensions. We then use the fit function
 \begin{equation}
     R(t) \sim\sqrt{E_1 \cos{(\omega t + \phi)\exp\left(-\tfrac{t}{\tau}\right)}+E_{\infty}},
 \end{equation}
where $R$ is the cloud width, $E_1$ is the oscillation amplitude, $\omega = 2\pi / \tau_{\rm T}$ is the Talbot frequency, $\phi$ is a phase shift, $\tau$ is the decay time related to the coherence length 
and $E_{\infty}$ is the long-time limit.  This allows us to relate the measured Talbot decay time to the coherence in the system.

We compare the results of the Talbot measurements to single-particle simulations based on exact diagonalization in one spatial dimension and thereby relate the measured decay time $\tau$ to the coherence length $\xi$ of the quantum state.

We start the Talbot simulations with a single-particle reduced density matrix $\rho_{x,x'}(t)$,
\begin{equation}
\label{eq:rho_xx'}
    \rho_{x,x'}(t=0) = \sum_{i,j} w_i(x) w_j^*(x') \sqrt{n_i n_j} \, \exp\left(-\lvert i-j\rvert\tfrac{\al}{\xi}\right) \lvert x \rangle \langle x' \rvert
\end{equation}
where $i,j$ are the indices of the lattice sites, $w_i(x)$ is the Wannier function of the lowest band localized at site $i$, and $n_i$ is the mean density of site $i$. 
This results in an exponential decay of the single-particle correlation function 
\begin{equation}
    g^{(1)} = \braket{{a}^\dagger_i a_j} \sim \exp{(-\lvert i-j \rvert \al/\xi)}
\end{equation}
between lattice sites $i$ and $j$. Note that Eq.~\ref{eq:rho_xx'} describes perfect coherence within one lattice site, even for very small $\xi$, and that also the situation where the coherence length $\xi$ is much larger than the system size $L$ is covered.

We then time-evolve $\rho_{x,x'}$ under the free single-particle propagator $U_t=e^{-iH_0t/\hbar}$ for a duration $t$ to obtain $\rho_{x,x'}(t)=U_t\rho_{x,x'}U^\dagger_t$ (Fig.~\ref{fig:SimTalbot}), and measure the energy $E$ of the final state in the lattice: $E(t)=\textrm{Tr}[\rho(t)H_{\ell\textbf{}}]$, where $H_{\ell\textbf{}}=p^2/(2m)+V \cos^2(\pi x/\al)$.

Our cloud is roughly circular with radius $8\al$ resulting in a (weighted) mean chain length of $13\al$. Starting with $L=13$ equally occupied sites (ground-band Wannier states, $V=10\,\Er$), we find Talbot oscillations (see Fig.~\ref{fig:SimTalbot}) similar in shape to the experimental ones (Fig.~\ref{fig:Figure3}). 
From the numerical data we extract the decay time $\tau$ with the model $E(t) = E_1 \cos{(\omega t + \phi)\exp{(-\frac{t}{\tau})}}+E_{\infty}$, where $E_1$ is the oscillation  amplitude, $\omega$ the Talbot frequency, $\phi$ a phase shift, and $E_\infty$ the long-time limit. Plotting $\tau$ versus $\xi$, we find the expected linear relationship $\tau/\tau_{\rm T} = \xi/(2\al)$ plus finite-size corrections for large coherence lengths. The  relation between the Talbot decay time $\tau$ and initial coherence length $\xi$ is well described by 
\begin{equation}
    \frac{\tau}{\tau_{\rm T}} = \left(\frac{2\al}{\xi}+0.22\right)^{-1}\label{eq:tau_xi}.
\end{equation} 
Experimentally, the thermalization in two dimensions after the one-dimensional Talbot evolution averages over the different chain lengths in our system. We numerically checked that explicitly adding the energy of the one-dimensional chains of different length results in the same behavior as Eq.~\ref{eq:tau_xi}. We have compared results with and without the trapping potential of the vertical lattice and found no significant effect over the timescales probed in the experiment.

\newpage

\bibliography{DisorderRef}

@article{Abanin2019,
  title = {Colloquium: Many-body localization, thermalization, and entanglement},
  author = {Abanin, Dmitry A. and Altman, Ehud and Bloch, Immanuel and Serbyn, Maksym},
  journal = {Rev. Mod. Phys.},
  volume = {91},
  pages = {021001},
  year = {2019},
  publisher = {American Physical Society},
  doi = {10.1103/RevModPhys.91.021001},
}

@article{Morrison2008,
doi = {10.1088/1367-2630/10/7/073032},
url = {https://dx.doi.org/10.1088/1367-2630/10/7/073032},
year = {2008},
month = {jul},
publisher = {},
volume = {10},
pages = {073032},
author = {S Morrison and A Kantian and A J Daley and H G Katzgraber and M Lewenstein and H P Büchler and P Zoller},
title = {Physical replicas and the {Bose} glass in cold atomic gases},
journal = {New J. Phys.}
}

@article{edwards1975,
doi = {10.1088/0305-4608/5/5/017},
url = {https://dx.doi.org/10.1088/0305-4608/5/5/017},
year = {1975},
month = {may},
publisher = {},
volume = {5},
pages = {965},
author = {S F Edwards and  P W Anderson},
title = {Theory of spin glasses},
journal = {J. Phys. F}
}

@article{Anderson1958,
  title = {Absence of Diffusion in Certain Random Lattices},
  author = {Anderson, P. W.},
  journal = {Phys. Rev.},
  volume = {109},
  pages = {1492--1505},
  year = {1958},
  publisher = {American Physical Society},
  doi = {10.1103/PhysRev.109.1492},
}

@article{roati2008anderson,
  title={Anderson localization of a non-interacting {Bose--Einstein} condensate},
  author={Roati, Giacomo and D’Errico, Chiara and Fallani, Leonardo and Fattori, Marco and Fort, Chiara and Zaccanti, Matteo and Modugno, Giovanni and Modugno, Michele and Inguscio, Massimo},
  journal={Nature},
  volume={453},
  pages={895--898},
  year={2008},
  publisher={Nature Publishing Group UK London},
doi = {10.1038/nature07071}
}

@article{jendrzejewski2012three,
  title={Three-dimensional localization of ultracold atoms in an optical disordered potential},
  author={Jendrzejewski, Fred and Bernard, Alain and Mueller, Killian and Cheinet, Patrick and Josse, Vincent and Piraud, Marie and Pezz{\'e}, Luca and Sanchez-Palencia, Laurent and Aspect, Alain and Bouyer, Philippe},
  journal={Nat. Phys.},
  volume={8},
  pages={398--403},
  year={2012},
  publisher={Nature Publishing Group UK London},
doi = {10.1038/nphys2256}
}

@article{white2020observation,
  title={Observation of two-dimensional {Anderson} localisation of ultracold atoms},
  author={White, Donald H and Haase, Thomas A and Brown, Dylan J and Hoogerland, Maarten D and Najafabadi, Mojdeh S and Helm, John L and Gies, Christopher and Schumayer, Daniel and Hutchinson, David AW},
  journal={Nat. Commun.},
  volume={11},
  pages={4942},
  year={2020},
  publisher={Nature Publishing Group UK London},
doi = {10.1038/s41467-020-18652-w}
}

@article{
Kondov2011,
author = {S. S. Kondov  and W. R. McGehee  and J. J. Zirbel  and B.  DeMarco },
title = {Three-Dimensional {Anderson} Localization of Ultracold Matter},
journal = {Science},
volume = {334},
pages = {66-68},
year = {2011},
doi = {10.1126/science.1209019},
}

@article{Lee2001,
  title = {Phase Diagram of a Disordered {Boson Hubbard} Model in Two Dimensions},
  author = {Lee, Ji-Woo and Cha, Min-Chul and Kim, Doochul},
  journal = {Phys. Rev. Lett.},
  volume = {87},
  pages = {247006},
  year = {2001},
  publisher = {American Physical Society},
  doi = {10.1103/PhysRevLett.87.247006},
}

@article{Fallani2007,
  title = {Ultracold Atoms in a Disordered Crystal of Light: Towards a {Bose} Glass},
  author = {Fallani, L. and Lye, J. E. and Guarrera, V. and Fort, C. and Inguscio, M.},
  journal = {Phys. Rev. Lett.},
  volume = {98},
  pages = {130404},
  year = {2007},
  publisher = {American Physical Society},
  doi = {10.1103/PhysRevLett.98.130404},
}

@article{aleiner2010finite,
  title={A finite-temperature phase transition for disordered weakly interacting bosons in one dimension},
  author={Aleiner, IL and Altshuler, BL and Shlyapnikov, GV},
  journal={Nat. Phys.},
  volume={6},
  pages={900--904},
  year={2010},
  publisher={Nature Publishing Group UK London},
doi = {10.1038/nphys1758}
}

@article{Binder1986,
  title = {Spin glasses: {Experimental} facts, theoretical concepts, and open questions},
  author = {Binder, K. and Young, A. P.},
  journal = {Rev. Mod. Phys.},
  volume = {58},
  pages = {801--976},
  year = {1986},
  publisher = {American Physical Society},
  doi = {10.1103/RevModPhys.58.801},
}

@article{Pollet2009,
  title = {Absence of a Direct {Superfluid to Mott} Insulator Transition in Disordered {Bose} Systems},
  author = {Pollet, L. and Prokof'ev, N. V. and Svistunov, B. V. and Troyer, M.},
  journal = {Phys. Rev. Lett.},
  volume = {103},
  pages = {140402},
  year = {2009},
  publisher = {American Physical Society},
  doi = {10.1103/PhysRevLett.103.140402},
}

@article{weichman2008,
  title={Dirty bosons: twenty years later},
  author={Weichman, Peter B},
  journal={Mod. Phys. Lett. B},
  volume={22},
  pages={2623--2647},
  year={2008},
  publisher={World Scientific},
doi = {10.1142/S0217984908017187},
}

@article{Thomson2016,
  title = {Measuring the {Edwards-Anderson} order parameter of the {Bose} glass: A quantum gas microscope approach},
  author = {Thomson, S. J. and Walker, L. S. and Harte, T. L. and Bruce, G. D.},
  journal = {Phys. Rev. A},
  volume = {94},
  pages = {051601},
  year = {2016},
  publisher = {American Physical Society},
  doi = {10.1103/PhysRevA.94.051601},
}

@phdthesis{Abadal2020Thesis,
    author = {Antonio Rubio Abadal},
    title = {Probing quantum thermalization and localization in {Bose-Hubbard} systems},
    school = {Ludwig-Maximilians-Universit{\"a}t M{\"u}nchen},
     year = {2020}
}

@article{schreiber2015observation,
  title={Observation of many-body localization of interacting fermions in a quasirandom optical lattice},
  author={Schreiber, Michael and Hodgman, Sean S and Bordia, Pranjal and L{\"u}schen, Henrik P and Fischer, Mark H and Vosk, Ronen and Altman, Ehud and Schneider, Ulrich and Bloch, Immanuel},
  journal={Science},
  volume={349},
  pages={842--845},
  year={2015},
  publisher={American Association for the Advancement of Science},
doi = {10.1126/science.aaa7432}
}

@article{Choi2016,
  author = { Choi, J. and Hild, S.  and  Zeiher, J.  and Schau{\ss}, P. and  Rubio-Abadal, A.  and Yefsah, T.  and  Khemani, V. and Huse, D. A.  and  Bloch, I.  and  Gross, C.},
  title = {Exploring the many-body localization transition in two dimensions},
  journal = {Science},
  volume = {352},
  pages = {1547--1552},
  year = {2016},
  doi = {10.1126/science.aaf8834},
}

@article{billy2008direct,
  title={Direct observation of {Anderson} localization of matter waves in a controlled disorder},
  author={Billy, Juliette and Josse, Vincent and Zuo, Zhanchun and Bernard, Alain and Hambrecht, Ben and Lugan, Pierre and Cl{\'e}ment, David and Sanchez-Palencia, Laurent and Bouyer, Philippe and Aspect, Alain},
  journal={Nature},
  volume={453},
  pages={891--894},
  year={2008},
  publisher={Nature Publishing Group},
doi = {10.1038/nature07000}
}

@article{Pasienski2010,
author = {Pasienski, M. and McKay, D. and White, M. and DeMarco, B.},
doi = {10.1038/nphys1726},
journal = {Nat. Phys.},
pages = {677--680},
publisher = {Nature Publishing Group},
title = {{A disordered insulator in an optical lattice}},
volume = {6},
year = {2010}
}

@article{sbroscia2020observing,
  title={Observing localization in a 2D quasicrystalline optical lattice},
  author={Sbroscia, Matteo and Viebahn, Konrad and Carter, Edward and Yu, Jr-Chiun and Gaunt, Alexander and Schneider, Ulrich},
  journal={Phys. Rev. Lett.},
  volume={125},
  pages={200604},
  year={2020},
  publisher={APS},
doi = {10.1103/PhysRevLett.125.200604}
}

@article{yu2024observing,
  title={Observing the two-dimensional {Bose} glass in an optical quasicrystal},
  author={Yu, Jr-Chiun and Bhave, Shaurya and Reeve, Lee and Song, Bo and Schneider, Ulrich},
  journal={Nature},
  volume={633},
  pages={338--343},
  year={2024},
  publisher={Nature Publishing Group UK London},
doi = {10.1038/s41586-024-07875-2

}
}

@article{Santra2017_Talbot,
  title={Measuring finite-range phase coherence in an optical lattice using {Talbot} interferometry},
  author={Santra, Bodhaditya and Baals, Christian and Labouvie, Ralf and Bhattacherjee, Aranya B. and Pelster, Axel and Ott, Herwig},
  journal={Nat. Commun.},
  volume={8},
  pages={15601},
  year={2017},
  publisher={},
doi = {10.1038/ncomms15601}
}

@article{Pal2019,
  title = {Enhancement of the {Bose} glass phase in the presence of an artificial gauge field},
  author = {Pal, Sukla and Bai, Rukmani and Bandyopadhyay, Soumik and Suthar, K. and Angom, D.},
  journal = {Phys. Rev. A},
  volume = {99},
  pages = {053610},
  year = {2019},
  publisher = {American Physical Society},
  doi = {10.1103/PhysRevA.99.053610},
}

@article{Gerbier2005,
  title = {Phase Coherence of an Atomic {Mott} Insulator},
  author = {Gerbier, Fabrice and Widera, Artur and F\"olling, Simon and Mandel, Olaf and Gericke, Tatjana and Bloch, Immanuel},
  journal = {Phys. Rev. Lett.},
  volume = {95},
  pages = {050404},
  year = {2005},
  publisher = {American Physical Society},
  doi = {10.1103/PhysRevLett.95.050404},
}

@article{pollet2013review,
  title={A review of {Monte Carlo} simulations for the {Bose--Hubbard} model with diagonal disorder},
  author={Pollet, Lode},
  journal={C.R. Phys.},
  volume={14},
  pages={712--724},
  year={2013},
doi = {10.1016/j.crhy.2013.08.005}
}

@article{krusin1994bose,
  title={Bose-glass melting in {YBaCuO} crystals with correlated disorder},
  author={Krusin-Elbaum, Lia and Civale, Leonardo and Blatter, Gianni and Marwick, AD and Holtzberg, Frederic and Feild, C},
  journal={Phys. Rev. Lett.},
  volume={72},
  pages={1914--1917},
   year={1994},
  publisher={APS},
doi = {10.1103/PhysRevLett.72.1914}
}

@article{ducry2020electro,
  title={Electro-thermal transport in disordered nanostructures: a modeling perspective},
  author={Ducry, Fabian and Aeschlimann, Jan and Luisier, Mathieu},
  journal={Nanoscale Adv.},
  volume={2},
  pages={2648--2667},
  year={2020},
  publisher={Royal Society of Chemistry},
doi = {10.1039/D0NA00168F}
}

@article{shapiro2012cold,
  title={Cold atoms in the presence of disorder},
  author={Shapiro, Boris},
  journal={J. Phys. A: Math. Theor.},
  volume={45},
  pages={143001},
  year={2012},
  publisher={IOP Publishing},
doi = {10.1088/1751-8113/45/14/143001}
}

@article{sanchez2010disordered,
  title={Disordered quantum gases under control},
  author={Sanchez-Palencia, Laurent and Lewenstein, Maciej},
  journal={Nat. Phys.},
  volume={6},
  pages={87--95},
  year={2010},
  publisher={Nature Publishing Group UK London},
doi = {10.1038/nphys1507}
}

@article{schwartz2007transport,
  title={Transport and {Anderson} localization in disordered two-dimensional photonic lattices},
  author={Schwartz, Tal and Bartal, Guy and Fishman, Shmuel and Segev, Mordechai},
  journal={Nature},
  volume={446},
  pages={52--55},
  year={2007},
  publisher={Nature Publishing Group UK London},
doi = {10.1038/nature05623

}
}

@article{lahini2008anderson,
  title={Anderson localization and nonlinearity in one-dimensional disordered photonic lattices},
  author={Lahini, Yoav and Avidan, Assaf and Pozzi, Francesca and Sorel, Marc and Morandotti, Roberto and Christodoulides, Demetrios N and Silberberg, Yaron},
  journal={Phys. Rev. Lett.},
  volume={100},
  pages={013906},
  year={2008},
  publisher={APS},
doi = {10.1103/PhysRevLett.100.013906}
}

@article{Fisher1989,
  title = {Boson localization and the superfluid-insulator transition},
  author = {Fisher, M. P. A. and Weichman, P. B. and Grinstein, G. and Fisher, D. S.},
  journal = {Phys. Rev. B},
  volume = {40},
  pages = {546--570},
  year = {1989},
  doi = {10.1103/PhysRevB.40.546},
}

@article{bertoli2018finite,
  title={Finite-temperature disordered bosons in two dimensions},
  author={Bertoli, Giulio and Michal, Vincent P and Altshuler, Boris L and Shlyapnikov, Georgy V},
  journal={Phys. Rev. Lett.},
  volume={121},
  pages={030403},
  year={2018},
  publisher={APS},
doi = {10.1103/PhysRevLett.121.030403}
}

@article{zhu2023thermodynamic,
  title={Thermodynamic phase diagram of two-dimensional bosons in a quasicrystal potential},
  author={Zhu, Zhaoxuan and Yao, Hepeng and Sanchez-Palencia, Laurent},
  journal={Phys. Rev. Lett.},
  volume={130},
  pages={220402},
  year={2023},
  publisher={APS},
doi = {10.1103/PhysRevLett.130.220402}
}

@article{Meldgin2016,
  title = {Probing the {Bose} glass–superfluid transition using quantum quenches of disorder},
  author = {Meldgin, C. and Ray, U. and Russ, P. and Chen, D. and Ceperley, D. and DeMarco, B.},
  journal = {Nat. Phys.},
  volume = {12},
  pages = {646-649},
  year = {2016},
  publisher = {},
  doi = {10.1038/nphys3695},
}

@article{DiCarli2024,
author = {Di Carli, Andrea and Parsonage, Christopher and La Rooij, Arthur and Koehn, Lennart and Ulm, Clemens and Duncan, Callum W. and Daley, Andrew J. and Haller, Elmar and Kuhr, Stefan.},
doi = {10.1038/s41467-023-44610-3},
issn = {},
journal = {Nat. Commun.},
pages = {474},
title = {{Commensurate and incommensurate 1D interacting quantum systems}},
volume = {15},
year = {2024}
}

@article{larooij2022,
  title={A comparative study of deconvolution techniques for quantum-gas microscope images},
  author={La Rooij, Arthur and Ulm, C and Haller, E and Kuhr, S},
  journal={New J. Phys.},
  volume={25},
  pages={083036},
  year={2023},
  publisher={IOP Publishing},
doi = {10.1088/1367-2630/aced65}
}

@article{PhysRevB.80.214519,
  title = {Phase diagram of the disordered {Bose-Hubbard} model},
  author = {Gurarie, V. and Pollet, L. and Prokof'ev, N. V. and Svistunov, B. V. and Troyer, M.},
  journal = {Phys. Rev. B},
  volume = {80},
  pages = {214519},
  year = {2009},
  publisher = {American Physical Society},
  doi = {10.1103/PhysRevB.80.214519},
}

@article{berthier2011theoretical,
  title = {Theoretical perspective on the glass transition and amorphous materials},
  author = {Berthier, Ludovic and Biroli, Giulio},
  journal = {Rev. Mod. Phys.},
  volume = {83},
  pages = {587--645},
  year = {2011},
  publisher = {American Physical Society},
  doi = {10.1103/RevModPhys.83.587},
}

@article{soyler2011phase,
  title={Phase diagram of the commensurate two-dimensional disordered {Bose-Hubbard} model},
  author={S{\"o}yler, {\c{S}}ebnem G{\"u}ne{\c{s}} and Kiselev, Mikhail and Prokof’ev, Nikolay V and Svistunov, Boris V},
  journal={Phys. Rev. Lett.},
  volume={107},
  pages={185301},
  year={2011},
  publisher={APS},
doi = {10.1103/PhysRevLett.107.185301}
}

@article{capogrosso2008monte,
  title={{Monte Carlo} study of the two-dimensional {Bose-Hubbard} model},
  author={Capogrosso-Sansone, Barbara and S{\"o}yler, {\c{S}}ebnem G{\"u}ne{\c{s}} and Prokof’ev, Nikolay and Svistunov, Boris},
  journal={Phys. Rev. A},
  volume={77},
  pages={015602},
  year={2008},
  publisher={APS},
doi = {10.1103/PhysRevA.77.015602}
}

@article{Sherson2010,
  author = {Sherson, J. F. and Weitenberg, C. and Endres, M. and Cheneau, M. and Bloch, I. and Kuhr, S.},
  journal = {Nature},
  pages = {68--72},
  title = {Single-atom-resolved fluorescence imaging of an atomic {Mott} insulator},
  volume = {467},
  year = {2010},
  doi = {10.1038/nature09378},
}

@article{Bakr2010,
author = {W. S. Bakr  and A. Peng  and M. E. Tai  and R. Ma  and J. Simon  and J. I. Gillen  and S. Fölling  and L. Pollet  and M. Greiner },
title = {Probing the {Superfluid–to–Mott Insulator} Transition at the Single-Atom Level},
journal = {Science},
volume = {329},
pages = {547-550},
year = {2010},
doi = {10.1126/science.1192368},
URL = {https://www.science.org/doi/abs/10.1126/science.1192368},
eprint = {https://www.science.org/doi/pdf/10.1126/science.1192368}}

@article{Manetsch2024,
      title={A tweezer array with 6100 highly coherent atomic qubits}, 
      author={Hannah J. Manetsch and Gyohei Nomura and Elie Bataille and Kon H. Leung and Xudong Lv and Manuel Endres},
      year={2024},
  journal={ Arxiv},
      eprint={2403.12021},
      archivePrefix={arXiv},
      primaryClass={quant-ph},
      url={https://arxiv.org/abs/2403.12021}, 
     doi={10.48550/arXiv.2403.12021}
}

@article{lin2011superfluid,
  title={Superfluid-insulator transition in the disordered two-dimensional {Bose-Hubbard} model},
  author={Lin, Fei and S{\o}rensen, Erik S and Ceperley, DM},
  journal={Phys. Rev. B.},
  volume={84},
  pages={094507},
  year={2011},
  publisher={APS},
doi={10.1103/PhysRevB.84.094507}
}

@article{tusch1993,
  title={Interplay between disorder and electron interactions in a $d=3$ site-disordered {Anderson-Hubbard} model: {A} numerical mean-field study},
  author={Tusch, Michael A and Logan, David E},
  journal={Phys. Rev. B},
  volume={48},
  pages={14843},
  year={1993},
  publisher={APS},
doi={10.1103/PhysRevB.48.14843}
}

@article{paredes2005,
  title={Fermionic atoms in optical superlattices},
  author={Paredes, Belen and Tejedor, C and Cirac, J Ignacio},
  journal={Phys. Rev. A},
  volume={71},
  pages={063608},
  year={2005},
  publisher={APS},
doi={10.1103/PhysRevA.71.063608}
}

@article{bauer2011alps,
  title={The {ALPS} project release 2.0: {Open} source software for strongly correlated systems},
  author={Bauer, Bela and Carr, LD and Evertz, Hans Gerd and Feiguin, Adrian and Freire, J and Fuchs, Sebastian and Gamper, Lukas and Gukelberger, Jan and Gull, Emanuel and Guertler, Siegfried and others},
  journal={J. Stat. Mech.},
  volume={2011},
  pages={P05001},
  year={2011},
  doi={10.1088/1742-5468/2011/05/P05001},
  publisher={IOP Publishing}
}

@article{albuquerque2007alps,
  title={The {ALPS} project release 1.3: {Open-source} software for strongly correlated systems},
  author={Albuquerque, A Fabricio and Alet, Fabien and Corboz, Philippe and Dayal, Prakash and Feiguin, Adrian and Fuchs, Sebastian and Gamper, L and Gull, Emanuel and G{\"u}rtler, S and Honecker, Andreas and others},
  journal={J. Magn. Magn. Mater.},
  volume={310},
  pages={1187--1193},
  year={2007},
  doi={10.1016/j.jmmm.2006.10.304},
  publisher={Elsevier}
}

@article{alet2005alps,
  title={The {ALPS} project: {Open} source software for strongly correlated systems},
  author={Alet, Fabien and Dayal, Prakash and Grzesik, Axel and Honecker, Andreas and K{\"o}rner, Mathias and L{\"a}uchli, Andreas and R. Manmana, S and P. McCulloch, I and Michel, F and M. Noack, R and others},
  journal={J. Phys. Soc. Jpn.},
  volume={74},
  pages={30--35},
  year={2005},
  doi={10.1143/JPSJS.74S.30},
  publisher={The Physical Society of Japan}
}

@article{PhysRevB.59.R14157,
  title = {Stochastic series expansion method with operator-loop update},
  author = {Sandvik, Anders W.},
  journal = {Phys. Rev. B},
  volume = {59},
  issue = {22},
  pages = {R14157--R14160},
  numpages = {0},
  year = {1999},
  month = {Jun},
  publisher = {American Physical Society},
  doi = {10.1103/PhysRevB.59.R14157},
  url = {https://link.aps.org/doi/10.1103/PhysRevB.59.R14157}
}

@article{PhysRevE.71.036706,
  title = {Generalized directed loop method for quantum {Monte Carlo} simulations},
  author = {Alet, Fabien and Wessel, Stefan and Troyer, Matthias},
  journal = {Phys. Rev. E},
  volume = {71},
  issue = {3},
  pages = {036706},
  numpages = {16},
  year = {2005},
  month = {Mar},
  publisher = {American Physical Society},
  doi = {10.1103/PhysRevE.71.036706},
  url = {https://link.aps.org/doi/10.1103/PhysRevE.71.036706}
}

@article{PhysRevE.70.056705,
  title = {Optimal {Monte Carlo} updating},
  author = {Pollet, Lode and Rombouts, Stefan M. A. and Van Houcke, Kris and Heyde, Kris},
  journal = {Phys. Rev. E},
  volume = {70},
  issue = {5},
  pages = {056705},
  numpages = {6},
  year = {2004},
  month = {Nov},
  publisher = {American Physical Society},
  doi = {10.1103/PhysRevE.70.056705},
  url = {https://link.aps.org/doi/10.1103/PhysRevE.70.056705}
}

@article{tromp2023puddle,
  title={Puddle formation and persistent gaps across the non-mean-field breakdown of superconductivity in overdoped {(Pb, Bi)$_2$Sr$_2$CuO$_{6+\delta}$}},
  author={Tromp, Willem O and Benschop, Tjerk and Ge, Jian-Feng and Battisti, Irene and Bastiaans, Koen M and Chatzopoulos, Damianos and Vervloet, Amber HM and Smit, Steef and van Heumen, Erik and Golden, Mark S and others},
  journal={Nat. Mat.},
  volume={22},
   pages={703--709},
  year={2023},
  publisher={Nature Publishing Group UK London},
 doi = {10.1038/s41563-023-01497-1},
  url = {https://www.nature.com/articles/s41563-023-01497-1}
}

@article{yu2012bose,
  title={Bose glass and {Mott} glass of quasiparticles in a doped quantum magnet},
  author={Yu, Rong and Yin, Liang and Sullivan, Neil S and Xia, JS and Huan, Chao and Paduan-Filho, Armando and Oliveira Jr, Nei F and Haas, Stephan and Steppke, Alexander and Miclea, Corneliu F and others},
  journal={Nature},
  volume={489},
   pages={379--384},
  year={2012},
  publisher={Nature Publishing Group UK London},
 doi = {10.1038/nature11406},
  url = {https://www.nature.com/articles/nature11406.pdf}
}

@article{malpuech2007bose,
  title={Bose glass and superfluid phases of cavity polaritons},
  author={Malpuech, G and Solnyshkov, DD and Ouerdane, H and Glazov, MM and Shelykh, I},
  journal={Phys. Rev. Lett.},
  volume={98},
  number={},
  pages={206402},
  year={2007},
  publisher={APS},
 doi = {10.1103/PhysRevLett.98.206402},
  url = {https://journals.aps.org/prl/abstract/10.1103/PhysRevLett.98.206402}
}

@article{carrasquilla2010characterization,
  title={Characterization of the {Bose-glass} phase in low-dimensional lattices},
  author={Carrasquilla, Juan and Becca, Federico and Trombettoni, Andrea and Fabrizio, Michele},
  journal={Phys. Rev. B},
  volume={81},
  number={},
  pages={195129},
  year={2010},
  publisher={APS},
  doi = {10.1103/PhysRevB.81.195129},

}

@article{gonzalez2019localisation,
  title={Localisation of weakly interacting bosons in two dimensions: disorder vs lattice geometry effects},
  author={Gonz{\'a}lez-Garc{\'\i}a, Luis A and Caballero-Ben{\'\i}tez, Santiago F and Paredes, Rosario},
  journal={Sci. Rep.},
  volume={9},
  number={},
  pages={11049},
  year={2019},
  publisher={Nature Publishing Group UK London},
doi = {10.1038/s41598-019-47279-1}
}

@article{mao2020disorder,
  title={Disorder effects in the two-dimensional {Lieb} lattice and its extensions},
  author={Mao, Xiaoyu and Liu, Jie and Zhong, Jianxin and R{\"o}mer, Rudolf A},
  journal={Physica E},
  volume={124},
  pages={114340},
  year={2020},
  publisher={Elsevier},
doi = {10.1016/j.physe.2020.114340}
}

@article{ristivojevic2014superfluid,
  title={{Superfluid/Bose-glass} transition in one dimension},
  author={Ristivojevic, Zoran and Petkovi{\'c}, Aleksandra and Le Doussal, Pierre and Giamarchi, Thierry},
  journal={Phys. Rev. B},
  volume={90},
  number={},
  pages={125144},
  year={2014},
  publisher={APS},
doi = {10.1103/PhysRevB.90.125144}
}

@article{leonard2023probing,
  title={Probing the onset of quantum avalanches in a many-body localized system},
  author={L{\'e}onard, Julian and Kim, Sooshin and Rispoli, Matthew and Lukin, Alexander and Schittko, Robert and Kwan, Joyce and Demler, Eugene and Sels, Dries and Greiner, Markus},
  journal={Nat. Phys.},
  volume={19},
  number={},
  pages={481--485},
  year={2023},
  publisher={Nature Publishing Group UK London},
doi = {10.1038/s41567-022-01887-3}
}

@article{duncan2024critical,
  title={Critical states and anomalous mobility edges in two-dimensional diagonal quasicrystals},
  author={Duncan, Callum W},
  journal={Phys. Rev. B},
  volume={109},
  number={},
  pages={014210},
  year={2024},
  publisher={APS},
doi = {PhysRevB.109.014210}
}

@article{szabo2020mixed,
  title={Mixed spectra and partially extended states in a two-dimensional quasiperiodic model},
  author={Szab{\'o}, Attila and Schneider, Ulrich},
  journal={Phys. Rev. B},
  volume={101},
  number={},
  pages={014205},
  year={2020},
  publisher={APS},
doi = {10.1103/PhysRevB.101.014205}
}

@article{rispoli2019quantum,
  title={Quantum critical behaviour at the many-body localization transition},
  author={Rispoli, Matthew and Lukin, Alexander and Schittko, Robert and Kim, Sooshin and Tai, M Eric and L{\'e}onard, Julian and Greiner, Markus},
  journal={Nature},
  volume={573},
  number={},
  pages={385--389},
  year={2019},
  publisher={Nature Publishing Group UK London},
doi = {10.1038/s41586-019-1527-2}
}

@article{nagler2022observing,
  title={Observing the loss and revival of long-range phase coherence through disorder quenches},
  author={Nagler, Benjamin and Barbosa, Sian and Koch, Jennifer and Orso, Giuliano and Widera, Artur},
  journal={Proc. Natl. Acad. Sci.},
  volume={119},
  number={},
  pages={e2111078118},
  year={2022},
  publisher={National Academy of Sciences},
doi={10.1073/pnas.2111078118}
}

@article{Sanpera2004spinglass,
  title={Atomic Fermi-Bose mixtures in inhomogeneous and random lattices: From {Fermi Glass} to {Quantum Spin Glass} and {Quantum Percolation}},
  author={Sanpera, A and Kantian, A. and Sanchez-Palencia, L. and  Zakrzewski, J. and Lewenstein, M.},
  journal={Phys. Rev. Lett.},
  volume={93},
  number={},
  pages={040401},
  year={2004},
  publisher={APS},
doi={10.1103/PhysRevLett.93.040401}
}

@article{Ahufinger2005spinglass,
  title={Disordered ultracold atomic gases in optical lattices: A case study of {Fermi-Bose} mixtures},
  author={Ahufinger, V. and Sanchez-Palencia, L. and Kantian, A. and  Sanpera, A. and Lewenstein, M.},
  journal={Phys. Rev. A},
  volume={72},
  number={},
  pages={063616},
  year={2005},
  publisher={APS},
doi={10.1103/PhysRevA.72.063616}
}

@article{ticea2025,
      title={Observation of disorder-induced superfluidity}, 
      author={Nicole Ticea and others},
      year={2025},
      journal = {arXiv:2512.21416},
      url={https://arxiv.org/abs/2512.21416}, 
}

@misc{Data,
    note = {The data used in this publication is openly available at https://doi.org/10.15129/af5cd941-2496-4a50-a10e-14df02036546}
}

\newpage
\newpage

\renewcommand{\thefigure}{S\arabic{figure}}
\renewcommand{\thetable}{S\arabic{table}}
\renewcommand{\theequation}{S\arabic{equation}}
\renewcommand{\thepage}{\arabic{page}}
\setcounter{figure}{0}
\setcounter{table}{0}
\setcounter{equation}{0}
\setcounter{subsection}{0}

\section*{Supplemental Materials}
\renewcommand\thesubsection{\arabic{subsection}}

\subsection{Calibration of disorder potential strength}
\begin{figure}[!h] \centering{}
    \includegraphics[width = 0.9\linewidth]{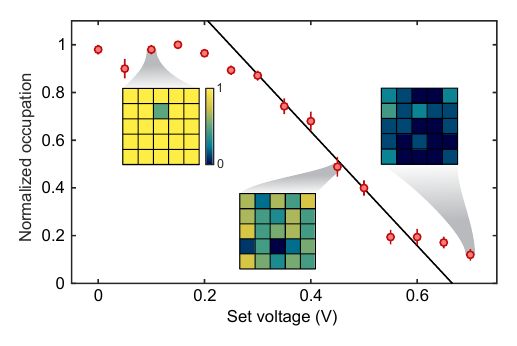}
    \vspace{-0.4cm}
    \caption[Calibration of light potentials]{ \label{fig:SI_calibration}\textbf{Calibration of light potentials.} Average occupations in the central $5\times5$ lattice sites of a Mott insulator with unit filling, after projecting a uniform square repulsive potential. Its intensity is proportional to a control voltage set in the control software. Each point contains data from eight images and shows the average over the $5\times5$ sites as pictured in the insets. The error bars are the standard error of the average occupation for the eight images. }
\vspace{-1cm}    
\end{figure}

The disorder potential  strength is calibrated by projecting a uniform square light potential onto a Mott insulator with unit occupation. We project this potential during the evaporation to a BEC and form the Mott insulator by increasing the horizontal lattices to  $V_{x,y}=14\Er$. We increase the power of the light potential until we observe empty sites in the center $5\times5$ region of the cloud. At this point, the strength of the light potential is equal to the energy offset ($\epsilon/h=\unit[300(40)]{Hz}$ at a radius of $10.5(5)$ lattice sites)
on the cloud edge, resulting from the harmonic potential
(Fig.~\ref{fig:SI_calibration}). We calculate the energy offset caused by the harmonic potential, $\epsilon(r)=\frac{1}{2} m w_r^2 r^2$ with an effective radial trapping frequency $w_r$, by taking into account the effect of the three orthogonal lattice beams. We measure the intensity of this light potential on a monitor camera and relate the camera reading to the strength of the light shift induced by the potential. When we generate a disorder pattern, we compare the maximum intensity in the pattern seen by the monitor camera to this calibration. 
As the amplitudes of the disorder pattern follow a uniform box distribution, the disorder strength is approximately the maximum amplitude in the pattern. We can therefore use the intensity on the monitor camera to determine the disorder strength of the pattern for a specific laser power. 

\subsection{Quantum Monte-Carlo simulations}

\begin{figure}[!b] \centering{}
    \includegraphics[width = 0.9\linewidth]{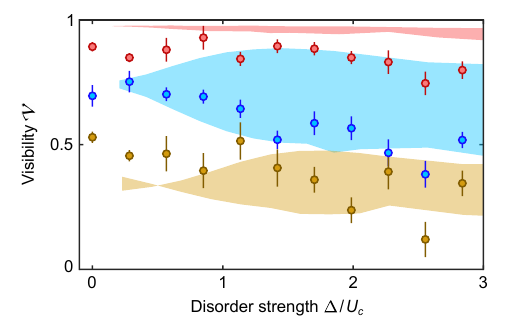}
    \caption[Visibility vs disorder strength]{ \label{SIfig:ExtendedData_Fig2_QMC}\textbf{Visibility vs disorder strength. } 
    Visibility vs disorder strength for three different lattice depths $V_{x,y}=7\Er$ (red), $11\Er$ (blue), and $13\Er$ (brown), same data as shown in Fig.~\ref{fig:Figure2}b.     
    Each datapoint is the mean of ten time-of-flight images and error bars indicate the standard error. The shaded areas indicate the results from QMC for densities between 0.5 (lower boundary) and 1 (upper boundary) particle per site.
    }
\label{fig:SI_extendedDataFig2_QMC}
\end{figure}
\label{sec:SIQMC}
To model the visibility measurements shown in Fig.~\ref{fig:Figure2},
we simulate the thermal state of the disordered Bose-Hubbard model (Eq. \ref{eq:BHM}) 
without harmonic confinement ($\epsilon_i = 0$) in the grand canonical ensemble at $T=0.1\,U/k_{B}$, using quantum Monte Carlo routines provided by the ALPS  library~\cite{alet2005alps,albuquerque2007alps,bauer2011alps}. We utilize the directed loop algorithm \cite{PhysRevB.59.R14157,PhysRevE.70.056705,PhysRevE.71.036706} for which we ensured results were equilibrated and converged. We first calculated global density and superfluid density utilizing periodic boundary conditions to guide the determination of experimental parameters and the convergence of results with the number of different disorder snapshots. We then simulated open boundary conditions of boxes of similar size to the experiment and calculated the global density across a range of chemical potential values for a selection of disorder patterns across a range of lattice depths. This enabled the generation of a lookup table to determine the approximate value of the chemical potential for unit filling of the lattice, as is obtained in the experiment. Utilizing this lookup table, we obtained the unit density ground states, including the single-particle Green's function, at parameter values of interest. With access to the single-particle Green's function, $G_{ij}= \langle a^\dagger_i a_j\rangle$, we can calculate all relevant observables such as the site resolved occupations $\langle a^\dagger_i a_i\rangle = G_{ii}$.

We are then able to compute the spectral function
\begin{equation}
    S(\mathbf{k}) = \sum_{i,j} \text{e}^{i\mathbf{k}.(\mathbf{r}_i-\mathbf{r}_j)}G_{ij}\, ,
\end{equation}
from which we can compute the visibility 
\begin{equation}
    \mathcal{V'} = \frac{S_\text{max}-S_{\text{min}}}{S_\text{max}+S_{\text{min}}}\, .
\end{equation}
$S_\text{max}$ can be inferred from a peak in the spectral function, e.g., at $\mathbf{k}=(0,2\pi)$ and $S_{\text{min}}$ is at, e.g.,~$\mathbf{k}=(\sqrt 2\pi,\sqrt 2\pi)$. To produce the results in Fig.~\ref{fig:SI_extendedDataFig2_QMC} and compare directly with the experiments, a lattice of $8\times 8$ sites is simulated with the chemical potential chosen so that these are between $0.5$ and $1$ particles per site giving the minimum and maximum values for the envelope shown. We see that this gives the same general trend as the experimental data, and we note that the precise value from the quantum Monte-Carlo results is dependent on the exact value of the temperature utilized. At unit filling the theoretical calculations show that the visibility increases with disorder strength in the Bose glass region, but as the filling is decreased this effect washes away and the visibility becomes monotonically decreasing with increasing disorder.

\subsection{Additional data: Edwards-Anderson parameter}
\begin{figure}[!ht] \centering{}
    \includegraphics[width = 0.9\linewidth]{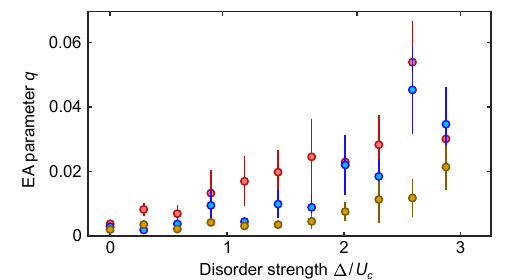}
    \caption[Extended data for Fig. 2.]{ \textbf{Extended data for Fig. 2.} 
    Edwards Anderson parameter, $q$, vs disorder strength for three different lattice depths $12\Er$ (red), $15\Er$ (blue), $18\Er$ (brown), as shown in  Fig.~\ref{fig:Figure2}c.    
    Each datapoint shows $q_i$ averaged over the central $5 \times 5$ lattice sites, using four different disorder patters with ten images each. The error bars are the standard error.    }
\label{fig:SI_ExtendedData_Fig2_EA}
\end{figure}

In Fig.~\ref{fig:SI_ExtendedData_Fig2_EA}, we show vertical cuts across the diagram presented in Fig.\,\ref{fig:Figure2}c, including error bars. All of the measurements of the Edwards-Anderson parameter, $q_i$, presented in this paper use four different disorder patterns. To justify this choice, we measured $q_i$ for different numbers of disorder patterns. We found that $q_i$ no longer increases when the number of patterns is increased from four to five.

\begin{figure}[!b] \centering
    \includegraphics[width = 0.9\linewidth]{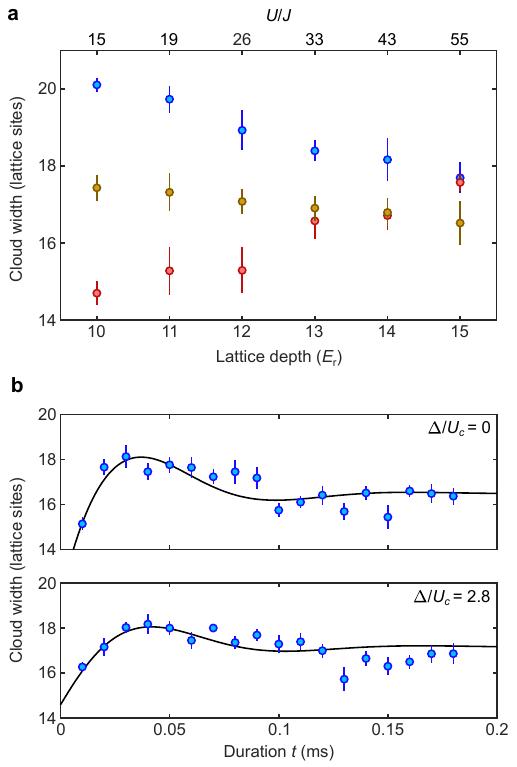}
    \caption[Talbot interferometry at the onset of the MI transition]{ \label{fig:MITalbot}\textbf{Talbot interferometry at the onset of the MI transition.} 
    \textbf{a} Cloud width after Talbot interferometry for $\Delta = 0$ vs. lattice depth, at $t=0.5\tau_{\rm T}$ (blue), $t =1\tau_{\rm T}$ (red), $t=1.5\tau_{\rm T}$ (brown). 
    Each data point is the mean of ten repetitions and the error bars are the standard error. \textbf{b}
    Talbot interferometry at $V_{x,y}=14\Er$ ($U/J = 43$) with and without disorder, $\Delta/U_c =0$ (top panel) and $\Delta/U_c = 2.8$  (bottom panel).The corresponding coherence lengths are $\xi = 0.6(2)$ sites and $\xi = 0.7(3)$ sites, respectively. Each data point is the mean of ten repetitions of the same experiment. 
    }
\label{fig:SI_extendedDataTalbot}
\end{figure}

\subsection{Additional data: Talbot interferometry}

We investigated how the phase coherence observed in the Talbot interferometry measurements is affected by the onset of the Mott-insulating phase in the absence of disorder. We observed a strong decrease of the contrast of the Talbot oscillations when increasing the lattice depth beyond the critical value ($V_{x,y} > 10.5\Er$, $U/J >16.7$), which we attribute to the growing Mott insulator in the center of the cloud (Fig.~\ref{fig:SI_extendedDataTalbot}a). Due to the harmonic confinement and the resulting spatially varying chemical potential, the outer region of the cloud remains superfluid. When adding disorder to the system that was initially in a Mott-insulating state, the Talbot interferometry signal shows no discernible difference to the case without disorder (Fig.~\ref{fig:SI_extendedDataTalbot}b), and the corresponding coherence length in both cases is smaller than a lattice site.

\subsection{Additional data: Entering the Bose-glass phase}
\begin{figure}[!h] \centering{}
    \includegraphics[width = 0.9\linewidth]{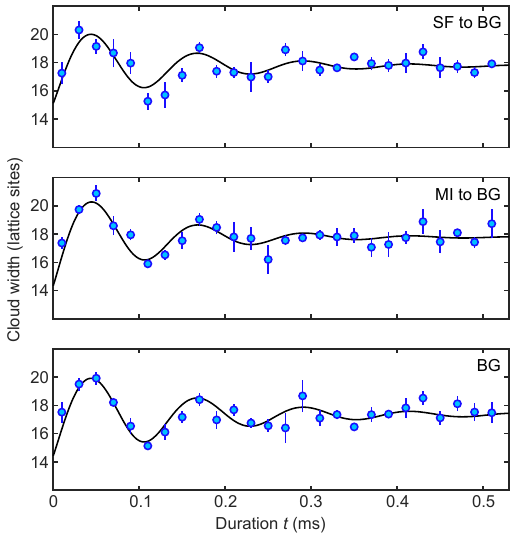} 
    \vspace{-0.1cm}
    \caption[Entering the Bose glass from the superfluid and Mott insulator]{\label{fig:tobg}\textbf{ Entering the Bose glass from the superfluid and Mott insulator.} Top panel: Talbot interferometry after preparing a superfluid, before entering the Bose glass. Middle panel: Starting in the Mott-insulating regime before entering the Bose glass. Bottom panel: Directly preparing a Bose glass. Each data point is the mean of five repetitions.
    The black lines show a fit with a damped sine, as in Fig.~\ref{fig:Figure3}.
    }
\label{fig:SI_EnteringTheBG}
\end{figure}

We have demonstrated that transitioning from the superfluid to the Bose glass and back causes a reduction in coherence length (Fig.~\ref{fig:Figure4}a). Similarly, starting from the Mott-insulating state and transitioning into the Bose glass and back does not restore the initial state (Fig.~\ref{fig:Figure4}c). 
We also investigate whether the coherence we observe in the Bose glass depends on whether the system was initially in the superfluid or MI state. To do this, we first prepare a superfluid at $V_{x,y}=7\Er$ and disorder strength $\Delta/U_c = 1.7$ and then change the lattice depth to $V_{x,y}=10\Er$, into the Bose-glass phase. The resulting Talbot interferometry measurement shows a coherence length $\xi=2.7(1.1)\al$ (Fig.~\ref{fig:SI_EnteringTheBG}, top panel). 
We compare this measurement to starting in the Mott-insulating regime ($V_{x,y}=18\Er$, $\Delta/U_c = 1.7$) and reducing the lattice depth ($V_{x,y}=10\Er$) to enter the Bose glass. We find a coherence length $\xi=2.3(6)\al$ (Fig.~\ref{fig:SI_EnteringTheBG}, middle panel). 
As a reference measurement, we directly prepare a Bose glass ($V_{x,y}=10\Er$,  $\Delta/U_c = 1.7$) showing a coherence length $\xi=3.3(9)\al$ (Fig.~\ref{fig:SI_EnteringTheBG}, bottom panel).
The coherence lengths overlap within the error bars, indicating that the preparation of the Bose glass is independent of the initial state within our experimental errors.

\end{document}